\documentclass[letterpaper, 10 pt, conference]{ieeeconf}
\IEEEoverridecommandlockouts                         
\usepackage{amsmath}
\usepackage{amssymb}
\usepackage{amsthm}
\newtheoremstyle{mystyle}%
  {\topsep}%
  {\topsep}%
  {\normalfont}%
  {}%
  {\bfseries}%
  {}%
  {.5em}%
  {}%
\theoremstyle{plain}

\usepackage{mathtools}
\usepackage{siunitx}

\usepackage{algorithm}
\usepackage{algpseudocode}

\usepackage{lipsum}

\usepackage[short,nocomma]{optidef}
\usepackage{xcolor}

\usepackage{tikz}
\usetikzlibrary{arrows.meta,positioning}
\usetikzlibrary{calc}
\usepackage{pgf}
\usepackage{pgfplots}
\pgfplotsset{compat=newest}

\everymath=\expandafter{\the\everymath\displaystyle}

\title{\LARGE \bf
Time-Series-Informed Closed-loop Learning for\\Sequential Decision Making and Control}

\author{Sebastian Hirt, Lukas Theiner, and Rolf Findeisen%
\thanks{The authors are with the Control and Cyber-Physical Systems Laboratory,
        Technical University of Darmstadt, Germany
        \{sebastian.hirt, lukas.theiner, rolf.findeisen\}@iat.tu-darmstadt.de}%
\thanks{This research was supported by the German Research Foundation (DFG) within RTG 2761 LokoAssist under grant no. 450821862.}
}

\begin{document}
\maketitle
\thispagestyle{empty}
\pagestyle{empty}

\begin{abstract}
Closed-loop performance of sequential decision making algorithms, such as model predictive control, depends strongly on the choice of controller parameters. Bayesian optimization allows learning of parameters from closed-loop experiments, but standard Bayesian optimization treats this as a black-box problem and ignores the temporal structure of closed-loop trajectories, leading to slow convergence and inefficient use of experimental resources.
We propose a time-series-informed multi-fidelity Bayesian optimization framework that aligns the fidelity dimension with closed-loop time, enabling intermediate performance evaluations within a closed-loop experiment to be incorporated as lower-fidelity observations. Additionally, we derive probabilistic early stopping criteria to terminate unpromising closed-loop experiments based on the surrogate model’s posterior belief, avoiding full episodes for poor parameterizations and thereby reducing resource usage.
Simulation results on a nonlinear control benchmark demonstrate that, compared to standard black-box Bayesian optimization  approaches, the proposed method achieves comparable closed-loop performance with roughly half the experimental resources, and yields better final performance when using the same resource budget, highlighting the value of exploiting temporal structure for sample-efficient closed-loop controller tuning.
\end{abstract}

\section{Introduction}
Sequential decision making algorithms are fundamental tools for controlling complex systems under uncertainty and constraints. Their closed-loop performance, however, depends critically on suitable parameter choices, e.g., cost function parameters shaping the desired behavior. Model predictive control (MPC) is a widely used model-based method \cite{rawlings2017model,findeisen2002introduction} and serves as a representative sequential decision making algorithm in this work. Learning-based methods have been increasingly explored to improve MPC performance under incomplete system knowledge or changing operating conditions \cite{Mesbah2022, hewing2020cautious,maiworm2021online, zieger2020towards,drgovna2025safe}. However, learning in model- and optimization-based control also raises concerns regarding safety and closed-loop performance \cite{Mesbah2022}, and accurate prediction models alone do not necessarily yield optimal closed-loop behavior \cite{kordabad2023reinforcement,gevers1993towards,drgovna2025safe}. 

\begin{figure}[h]
    \centering
    \vspace{2mm}
    \scalebox{0.8}{\begin{tikzpicture}[x=10cm,y=4.5cm,>=Stealth, line cap=round]

\def\sstop{0.667}   %
\def\Gbest{-0.16}  %
\def\L{6}          %

\def\bandscale{4}    %
\def\bandoffset{0.3} %

\pgfmathdeclarefunction{mmean}{1}{%
  \pgfmathparse{-0.90 + 0.12*pow(#1,0.8)}%
}
\pgfmathdeclarefunction{sigbase}{1}{%
  \pgfmathparse{0.030 + 0.020*exp(-100*#1) + 0.05*#1}%
}

\fill[gray!50] (\sstop,0.0) rectangle (1.05,-0.94);

\newcommand{\PlotFunc}[3]{%
  \draw[#1] plot[smooth] coordinates {
    (0.00,{#2(0.00)}) (0.05,{#2(0.05)}) (0.10,{#2(0.10)}) (0.15,{#2(0.15)})
    (0.20,{#2(0.20)}) (0.25,{#2(0.25)}) (0.30,{#2(0.30)}) (0.35,{#2(0.35)})
    (0.40,{#2(0.40)}) (0.45,{#2(0.45)}) (0.50,{#2(0.50)}) (0.55,{#2(0.55)})
    (0.60,{#2(0.60)}) (0.65,{#2(0.65)}) (0.70,{#2(0.70)}) (0.75,{#2(0.75)})
    (0.80,{#2(0.80)}) (0.85,{#2(0.85)}) (0.90,{#2(0.90)}) (0.95,{#2(0.95)})
    (1.00,{#2(1.00)})
  };
}

\pgfmathdeclarefunction{upperf}{1}{%
  \pgfmathparse{ mmean(#1) + (\bandoffset + \bandscale*sigbase(#1)) }%
}
\pgfmathdeclarefunction{lowerf}{1}{%
  \pgfmathparse{ mmean(#1) - (\bandscale*sigbase(#1) - \bandoffset) }%
}

\begin{scope}
  \fill[teal!18]
    plot[smooth] coordinates {
      (0.00,{upperf(0.00)}) (0.05,{upperf(0.05)}) (0.10,{upperf(0.10)})
      (0.15,{upperf(0.15)}) (0.20,{upperf(0.20)}) (0.25,{upperf(0.25)})
      (0.30,{upperf(0.30)}) (0.35,{upperf(0.35)}) (0.40,{upperf(0.40)})
      (0.45,{upperf(0.45)}) (0.50,{upperf(0.50)}) (0.55,{upperf(0.55)})
      (0.60,{upperf(0.60)}) (0.65,{upperf(0.65)}) (0.70,{upperf(0.70)})
      (0.75,{upperf(0.75)}) (0.80,{upperf(0.80)}) (0.85,{upperf(0.85)})
      (0.90,{upperf(0.90)}) (0.95,{upperf(0.95)}) (1.00,{upperf(1.00)})
    } --
    plot[smooth] coordinates {
      (1.00,{lowerf(1.00)}) (0.95,{lowerf(0.95)}) (0.90,{lowerf(0.90)})
      (0.85,{lowerf(0.85)}) (0.80,{lowerf(0.80)}) (0.75,{lowerf(0.75)})
      (0.70,{lowerf(0.70)}) (0.65,{lowerf(0.65)}) (0.60,{lowerf(0.60)})
      (0.55,{lowerf(0.55)}) (0.50,{lowerf(0.50)}) (0.45,{lowerf(0.45)})
      (0.40,{lowerf(0.40)}) (0.35,{lowerf(0.35)}) (0.30,{lowerf(0.30)})
      (0.25,{lowerf(0.25)}) (0.20,{lowerf(0.20)}) (0.15,{lowerf(0.15)})
      (0.10,{lowerf(0.10)}) (0.05,{lowerf(0.05)}) (0.00,{lowerf(0.00)})
    } -- cycle;

  \PlotFunc{ultra thick, teal}{upperf}{upperf}
\end{scope}

\node[teal!70!black, anchor=south west]
  at (0.25,{upperf(0.1)+0.09})
  {$\mathrm{UCB}(\theta_n,s)$};

\foreach \l in {1,...,4} {
  \pgfmathsetmacro{\s}{\l/\L}
  \pgfmathsetmacro{\y}{-0.65 + 0.2*(\s)^(0.8)} %
  \node[circle, fill=teal!80!black, inner sep=0pt, minimum size=7pt] at (\s,\y) {};

  \ifnum\l=1
      \node[above] at (\s,\y+0.02) {\scriptsize $\bar G(\theta_n,1/L)$};
    \fi
    \ifnum\l=3
      \node[above] at (\s,\y+0.02) {\scriptsize $\bar G(\theta_n,3/L)$};
    \fi
}

\fill[gray!12, opacity=0.6] (\sstop,0.0) rectangle (1.05,-0.94);
\node[red, anchor=west] at (\sstop, -0.67) {Early stop};
\draw[red, dashed, ultra thick] (\sstop,0.0) -- (\sstop,-0.94);

\draw[violet!80!black, dashed, ultra thick] (0,\Gbest) -- (0.99,\Gbest)
  node[pos=0, right, violet!80!black, xshift=3mm, yshift=3mm] {$G_n^*$};

\draw[very thick,->] (0,0) -- (1.05,0)
  node[above, xshift=-1.4cm, yshift=0.0cm] {$s$ (fidelity/time)};

\draw[very thick,->] (0,0) -- (0,-0.92)
  node[right, yshift=0.3cm, xshift=0.1cm] {$\hat{\bar G}(\theta_n,s)$};

\begin{scope}[shift={(1.0,{upperf(1.0)})}, x=2pt, y=2pt]
  \draw[red, line width=3pt] (-3,-3) -- (3,3);
  \draw[red, line width=3pt] (-3,3) -- (3,-3);
\end{scope}

\node[red, anchor=south east]
  at ($(1.005,{upperf(1.0)})+(-0.02,0.0)$)
  {$\mathrm{UCB}(\theta_n,1)$};

\draw[->, black, ultra thick, line cap=round]
  (\sstop,{-0.35}) .. controls ($(0.98,-0.48)$) .. (1.0,{upperf(1.0)+0.005});

\end{tikzpicture}}
    \vspace{-5mm}
    \caption{Proposed approach: The fidelity dimension $s$ of the Bayesian optimization surrogate $\bar G$ is aligned with closed-loop time, enabling intermediate closed-loop evaluations $\bar G(\theta_n,s)$ to be incorporated into the optimization process. Early termination is decided at the current iteration by predicting the upper confidence bound at the target fidelity ($\mathrm{UCB}(\theta_n,1)$), which corresponds to the total cost of the full closed-loop experiment, and comparing it to the best observed cost $G_n^*$. This allows efficient use of experimental resources.}
    \label{fig:approach}
    \vspace{-6mm}
\end{figure}

To this end, hierarchical closed-loop learning frameworks based on Bayesian optimization (BO) have been explored to directly optimize closed-loop performance \cite{paulson2023tutorial, piga2019performance, hirt2024safe, hirt2023stability}. These approaches typically separate global parameter learning from low-level control execution. As a sample-efficient optimizer, BO is particularly appealing here because evaluating controller parameterizations typically requires executing full closed-loop experiments, which is resource-intensive in practice, e.g., high-fidelity simulation, real hardware time, energy consumption, or wear. 
However, existing BO-based methods generally model the closed-loop learning problem as a black-box function, mapping from controller parameters to their performance in a full-length closed-loop experiment. As a result, informative partial trajectory data collected early in an episode remains unused, while every experiment is still executed until completion even when it is already clear that the parameterization performs poorly — leading to unnecessary resource consumption and slow convergence.
While multi-fidelity BO \cite{wu2020practical} provides means of including additional information sources, it is usually applied for incorporating artificial fidelity abstractions unrelated to the temporal structure of closed-loop experiments, such as lower-fidelity simulations.

Complementary to the approach proposed in this work, our recent work on Hierarchical Bayesian Optimization \cite{hirt2025hierarchical} exploits structure in the closed-loop dynamics to enable transfer across tasks. In this work, we instead focus on structure along the closed-loop time axis within a single task, aiming to improve resource efficiency during learning.
Related work such as \cite{stenger2024early} focuses on terminating episodes early and introduces heuristic mechanisms to convert truncated trajectories into surrogate-compatible BO updates. In contrast, we model partial episode evaluations directly as lower-fidelity observations within a multi-fidelity surrogate aligned with closed-loop time. This enables principled probabilistic early stopping from the GP belief itself, rather than relying on heuristic reconstruction of missing trajectory segments.

We propose a practical, sample-efficient closed-loop tuning workflow that explicitly exploits the temporal structure within closed-loop experiments. Instead of treating time as a generic contextual variable, we reinterpret the fidelity dimension in multi-fidelity BO as aligned with closed-loop time, which enables direct incorporation of partial episode information and principled mid-episode termination decisions. Specifically, our contributions are:
\begin{itemize}
    \item Time-series fidelity modeling: We introduce a time-series-informed Bayesian optimization (TSI-BO) formulation that aligns the surrogate model's fidelity dimension with the closed-loop time axis, such that intermediate partial-episode performance evaluations become lower-fidelity observations of the same closed-loop objective.
    \item Probabilistic early stopping rules: We derive GP-belief-based decision criteria (UCB and EI) that estimate the expected final episode performance already during the episode, enabling termination of unpromising parameterizations well before the episode end, avoiding unnecessary interaction time with the system and saving resources.
    \item Convergence-based termination: We additionally propose a lightweight convergence-based stopping criterion directly on the closed-loop state evolution to detect early convergence to the desired state.
\end{itemize}

In simulation, we show that the resulting time-series-informed BO framework substantially improves sample efficiency and achieves superior performance compared to standard single-fidelity BO baselines in terms of convergence speed, resource efficiency, and final closed-loop performance.

The remainder of this work is structured as follows.
We present the control task in Section \ref{sec:fundamentals} and recall fundamental concepts.
We introduce the proposed time-series-informed BO approach in Section \ref{sec:main}.
In Section \ref{sec:simulation} we showcase simulation results, and conclude in Section \ref{sec:conclusion}.

\section{Fundamentals}
\label{sec:fundamentals}
This section presents an overview of the control task and establishes the fundamentals.
We present the control objective and introduce parameterized model predictive control.
Afterward, we recall the fundamentals of Gaussian process regression and multi-fidelity Bayesian optimization.

\subsection{Problem Formulation}
We consider a nonlinear, discrete-time dynamical system
\begin{equation}
    \label{eqn:discrete_system_general}
    x_{k+1} = f(x_k, u_k),
\end{equation}
with states $x_k \in \mathbb{R}^{n_\mathrm{x}}$, inputs $u_k \in \mathbb{R}^{n_\mathrm{u}}$, nonlinear dynamics $f: \mathbb{R}^{n_\mathrm{x}} \times \mathbb{R}^{n_\mathrm{u}} \rightarrow \mathbb{R}^{n_\mathrm{x}}$, and discrete time index $k \in \mathbb{N}_0$. The control objective is to steer the system from an initial state $x_0$ toward a desired target $(x_\mathrm{d}, u_\mathrm{d})$.

The closed-loop behavior is determined by a parameterized control policy
\begin{equation}
    \pi : \mathbb{R}^{n_\mathrm{x}} \times \Theta \rightarrow \mathbb{R}^{n_\mathrm{u}}, \quad (x_k, \theta) \mapsto \pi(x_k;\theta),
\end{equation}
which maps the current state and the controller parameters $\theta \in \Theta \subset \mathbb{R}^{n_\mathrm{p}}, n_{\text{p}} \in \mathbb{N}$ to the applied input $u_k$.

To evaluate the quality of the closed-loop behavior induced by $\theta$, we define a closed-loop cost
\begin{equation}
    \label{eq:closed-loop_cost_general}
    G:\Theta \rightarrow \mathbb{R}, \quad \theta \mapsto G(\theta),
\end{equation}
obtained from a closed-loop experiment of finite length. The cost $G(\theta)$ aggregates closed-loop performance over time, yielding a scalar performance measure that quantifies the effect of the chosen parameters $\theta$ on the closed-loop behavior.
Consequently, the objective in this work is to find controller parameters $\theta$ that yield the best closed-loop performance according to $G(\theta)$. In the following, we consider a parameterized model predictive controller as a specific instantiation of $\pi$.

\subsection{Parameterized Model Predictive Control}
We consider a parameterized model predictive control (MPC) formulation with parameters $\theta$.
At every discrete time index $k$, given a set of parameters $\theta$ and the measurement of the current state $x_k$, we solve the parameterized optimal control problem given by
\color{black}
\begin{mini!}
    {\mathbf{\hat{u}}_k}{\left\{ \sum_{i=0}^{N-1} l_\theta({\hat x}_{i \mid k}, {\hat u}_{i \mid k}) + E_\theta({\hat x}_{N \mid k}) \! \right\}\label{eqn:mpc_ocp_cost}}{\label{eqn:mpc_ocp}}{}
    \addConstraint{\forall i}{\in \{0, 1, \dots, N-1\}: \notag}{}
    \addConstraint{}{\hat x_{i+1\mid k} = \hat f_\theta(\hat x_{i\mid k}, \hat u_{i\mid k}), \ \hat x_{0 \mid k} = x_k,}{\label{eqn:mpc_ocp_model}}
    \addConstraint{}{\hat x_{i \mid k} \in \mathcal{X}_\theta, \ {\hat u}_{i \mid k} \in \mathcal{U}, \ \hat x_{N \mid k} \in \mathcal{E}_\theta.}{\label{eqn:mpc_ocp_constraints}}
\end{mini!}
Here, $\hat{\cdot}_{i\mid k}$ denotes the model-based $i$-step ahead prediction at time index $k$.
$\hat{f}_\theta : \mathbb{R}^{n_\text{x}} \times \mathbb{R}^{n_\text{u}} \to \mathbb{R}^{n_\text{x}}, \quad (x, u) \mapsto \hat{f}_\theta(x, u)$ is the (parameterized) prediction model.
The length of the prediction horizon is $N \in \mathbb{N}$, $N < \infty$ and $l_\theta : \mathbb{R}^{n_\text{x}} \times \mathbb{R}^{n_\text{u}} \to \mathbb{R}, \quad (x, u) \mapsto l_\theta(x, u)$ and $E_\theta : \mathbb{R}^{n_\text{x}} \to \mathbb{R}, \quad x \mapsto E_\theta(x)$ are the (parameterized) stage and terminal cost functions, respectively.
The constraints \eqref{eqn:mpc_ocp_constraints} are comprised of the (parameterized) state, input, and terminal sets $\mathcal{X}_\theta \subset \mathbb{R}^{n_\text{x}}$, $\mathcal{U} \subset \mathbb{R}^{n_\text{u}}$, and $\mathcal{E}_\theta \subset \mathbb{R}^{n_\text{x}}$, respectively.
Minimizing the cost over the control input sequence $\mathbf{\hat{u}}_k$ results in the optimal input sequence $\mathbf{\hat{u}}_k^*(x_k; \theta)=[\hat u_{0 \mid k}^*(x_k; \theta),\dots,\hat u_{N-1 \mid k}^*(x_k; \theta)]$, of which only the first element is applied to system \eqref{eqn:discrete_system_general}. 
Subsequently, the optimal control problem \eqref{eqn:mpc_ocp} is solved again at all following time indices $k$.
Consequently, the parameterized control policy is given by $u_k = \pi(x_k;\theta) = \hat u_{0 \mid k}^*(x_k; \theta)$.

To optimize the control policy parameters towards a desired closed-loop performance, we exploit a multi-fidelity Bayesian optimization algorithm based on Gaussian process surrogate models, which we briefly introduce in the following.

\subsection{Gaussian Process Surrogate Models}\label{sec:gp}
\newcommand{\nd}{\ensuremath {n_\text{d}}}
To enhance the closed-loop performance of system \eqref{eqn:discrete_system_general} under the MPC law \eqref{eqn:mpc_ocp}, Bayesian optimization utilizes a surrogate model that captures the mapping between the parameters $\theta$ and the closed-loop performance measure $G(\theta)$.
We construct this surrogate using Gaussian process (GP) regression \cite{rasmussen2006gaussian}, which allows inferring a probabilistic model of an unknown function $\varphi: \mathbb{R}^{n_\xi} \to \mathbb{R}, \xi \mapsto \varphi(\xi)$ from data.

A GP $g(\xi) \sim \mathcal{GP}(m(\xi), k(\xi, \xi^\prime))$ is a random process, for which the outcomes $g(\xi_i)$ at any finite collection of inputs $\xi_i$ are jointly normally distributed.
It is fully defined by its prior mean function $m: \mathbb{R}^{n_\xi} \to \mathbb{R}, \xi \mapsto \mathbb{E}[g(\xi)]$ and the prior covariance function $k: \mathbb{R}^{n_\xi} \times \mathbb{R}^{n_\xi} \to \mathbb{R}, (\xi, \xi') \mapsto \mathrm{Cov}[g(\xi), g(\xi')]$.

To obtain a prediction model of $\varphi$, training data are incorporated into the model via Bayesian inference. 
The training dataset $\mathcal{D} = \{ (\xi_i, \gamma_i = \varphi(\xi_i) + \varepsilon_i) \mid i \in \{1, \dots, n_{\text{d}} \}, \varepsilon_i \sim \mathcal{N}(0, \sigma^2) \}$ is comprised of a set of $n_\text{d} \in \mathbb{R}$ noisy observations of the value of the function $\varphi$ at inputs $\xi_i$, where $\varepsilon_i$ is white Gaussian noise with variance $\sigma^2$.

Predictions of the unknown function value at an arbitrary input $\xi_*$ are given by the posterior distribution $g(\xi_*) \mid (\mathcal{D}, \xi_*) \sim \mathcal{N}(m^+(\xi_*), k^+(\xi_*))$ with mean and variance given by
\begin{subequations}
\label{eqn:posterior_gp}
\begin{align}
    m^+(\xi_*) &= m(\xi_*) + k_*^\top (K + \sigma^2 I)^{-1} \Gamma, \label{eq:gp_postMean} \\
    k^+(\xi_*) &= k(\xi_*, \xi_*) - k_*^\top (K + \sigma^2 I)^{-1} k_*. \label{eq:gp_postVar}
\end{align}
\end{subequations}
Here, $k_* \in \mathbb{R}^{\nd \times 1}$ with rows $[k_*]_i = k(\xi_i,\xi_*)$, $\Gamma \in \mathbb{R}^{\nd \times 1}$ with rows $[\Gamma]_i = \gamma_i - m(\xi_i)$, $K \in \mathbb{R}^{\nd \times \nd}$ with elements $[K]_{ij} = k(\xi_i, \xi_j)$, and $I \in \mathbb{R}^{\nd \times \nd}$ is the identity matrix.
The estimate of the unknown function value is given by the posterior mean \eqref{eq:gp_postMean} while the posterior variance \eqref{eq:gp_postVar} quantifies the uncertainty.
The prior mean and covariance function are chosen as part of the model design and typically include free hyperparameters.
The latter are often optimized through evidence maximization using the training data $\mathcal{D}$, see \cite{rasmussen2006gaussian}.

\subsection{Trace-aware Multi-fidelity Bayesian Optimization}
\label{sec:trace_aware_bo}
Bayesian optimization (BO) is a sequential framework for sample-efficient optimization of expensive-to-evaluate black-box functions \cite{garnett2023bayesian}.
In our setting, the closed-loop cost \eqref{eq:closed-loop_cost_general} is not available in closed form and can only be obtained by running full closed-loop experiments.
We therefore employ BO to efficiently explore the parameter space $\Theta$ and use closed-loop experiments to guide the improvement of the parametrized controller.
Specifically, the BO optimization problem is given by
\begin{align}
\begin{split}
    \label{eqn:bo_optimization_problem}
    \theta^* &= \arg \max_{\theta \in \Theta} \left\{G(\theta)\right\} \\
\end{split}
\end{align}
where $G$ is the closed-loop performance measure \eqref{eq:closed-loop_cost_general}.

When using BO to optimize the performance measure of an evolving process, such as closed-loop experiments of a dynamical system \eqref{eqn:discrete_system_general} controlled by an MPC \eqref{eqn:mpc_ocp}, aggregating the closed-loop performance into a single value upon completion results in a loss of valuable information about the dynamical time-series behavior.
While the objective remains to optimize the performance measure of the complete process, we aim to incorporate performance data from early-on in the closed-loop experiments as additional low-fidelity data points, utilizing trace-aware multi-fidelity BO \cite{wu2020practical}.

To this end, we introduce an additional fidelity dimension into the BO procedure and define the function $\bar G : \Theta \times [0, 1] \to \mathbb{R}, (\theta, s) \mapsto \bar G(\theta, s)$, where $s$ is the fidelity parameter. 
We define $s=1$, as the target fidelity, with all $s < 1$ being low-fidelity samples.
Since, ultimately, we are interested in optimizing the original problem \eqref{eqn:bo_optimization_problem} at the target fidelity, the multi-fidelity BO problem is given by
\begin{equation}
    \label{eqn:multi_fidelity_bo_optimization_problem}
    \theta^* = \arg \max_{\theta \in \Theta} \left\{\bar G(\theta, s=1)\right\}.
\end{equation}

Conveniently, in the present setting, a trace of low-fidelity data is additionally obtained when evaluating the performance at the highest fidelity without additional cost.
In the following, we denote the set of fidelities in one trace observation as $\mathcal{S}$, and the set of observations of the performance measures for a single parameter value $\theta$ at all fidelities in $\mathcal{S}$ as $\mathcal{Y}(\theta,\mathcal{S}) \coloneqq \{\bar G(\theta,s) \mid s\in\mathcal{S}\}$.

To solve \eqref{eqn:multi_fidelity_bo_optimization_problem}, GP regression models $\hat{\bar{G}}(\theta, s)$ of the unknown black box function $\bar G(\theta, s)$ are commonly employed.
During the optimization procedure, the GP surrogate model is sequentially updated using data obtained from evaluating the performance at a new parameter $\theta$ with different fidelity levels $\mathcal{S}$, leading to the sequential BO procedure.
Particularly, in each BO iteration $n \in \mathbb{N}$, we
\begin{enumerate}
    \item[1)] select the next parameter value of interest $\theta_n$, then evaluate the performance $\bar G(\theta_n,1)$ and record the trace of performance values to generate a new data point $\{ (\theta_n, \mathcal{S}_n), \mathcal{Y}(\theta_n, \mathcal{S}_n) \}$,
    \item[2)] update the training data set with the newly observed data point according to $\mathcal{D}_{n+1} \leftarrow \mathcal{D}_n \cup \{ (\theta_n, \mathcal{S}_n), \mathcal{Y}(\theta_n, \mathcal{S}_n) \}$,
    \item[3)] update the (posterior) GP model based on $\mathcal{D}_{n+1}$, including optimization of the hyperparameters.
\end{enumerate}
Here, $\mathcal{D}_n$ denotes the training data set, comprising data points observed up to iteration $n$.
To conduct the sequential learning procedure in a structured way, an acquisition function is employed to guide the selection of the next parameter value of interest in step $1)$.
We utilize the \textit{trace-aware knowledge gradient} ($\mathrm{taKG}$) acquisition function proposed in \cite{wu2020practical}, as it also incorporates the utility from the trace observations obtained when sampling $\theta$.
Similar to \cite{wu2020practical}, we limit complexity by restricting trace observations to a priori defined fidelities and keep a constant $\mathcal{S}$.

Before presenting the acquisition function, we recall the expected loss function $L_n(\theta, \mathcal{S})$ from \cite{wu2020practical}.
$L_n(\theta, \mathcal{S})$ represents the expected optimal value of the loss $-\bar G(\theta, s)$ at the target fidelity $s=1$, given that it is evaluated at a parameter value $\theta$ and a set of fidelities $\mathcal{S}$.
The expected loss is given by
\begin{align}
    L_n(\theta, \mathcal{S}) = \mathbb{E}_n \left[ \min_{\theta' \in \Theta} \mathbb{E}_n[-\hat{\bar{G}}(\theta', 1) \mid \mathcal{Y}(\theta, \mathcal{S})] \right],
\end{align}
where \( \mathbb{E}_n \) is the expectation taken with respect to the posterior distribution of $\hat{\bar{G}}$ at BO iteration $n$.

On this basis, the $\mathrm{taKG}$ acquisition function evaluates the utility of an observation at a given parameter set and set of fidelities.
It is given by \cite{wu2020practical}
\begin{align}
    \label{eqn:takg_acquisition}
    \mathrm{taKG}_n(\theta, \mathcal{S}) = \frac{L_n(\emptyset) - L_n(\theta, \mathcal{S})}{\text{cost}_n(\theta, \max \mathcal{S})},
\end{align}
where $L_n(\emptyset)$ represents the expected loss without additional observations, and $L_n(\theta, s)$ quantifies the expected loss after observing $\theta$ at all fidelities in $\mathcal{S}$.
Consequently, \eqref{eqn:takg_acquisition} expresses the one-step gain in the global reward, which is a standard approach for all knowledge-gradient-based acquisition functions \cite{garnett2023bayesian}.
The denominator $\text{cost}_n(\theta, \mathcal{S})$ accounts for the computational cost of evaluating $\bar G$ at $\theta$ and the highest fidelity in $\mathcal{S}$.
This way, evaluations prioritize observations that maximize information gain relative to their cost.

At each BO iteration $n$, the acquisition function guides the selection of $\theta_{n+1}$ by solving
\begin{align}
    \label{eqn:takg_optimization}
    \theta_{n+1} = \arg\max_{\theta \in \Theta} \mathrm{taKG}_n(\theta, \mathcal{S}).
\end{align}

By balancing information gain and cost, the trace-aware knowledge gradient enables efficient optimization of the multi-fidelity BO problem, while avoiding redundant evaluations at low fidelities.
Building on these concepts, we now introduce a novel approach that leverages the time-series structure of closed-loop experiments and integrates multi-fidelity information into the Bayesian optimization framework to enhance resource efficiency and performance.

\section{Efficient, Resource-aware Closed-loop Learning}
\label{sec:main}
In this section, we present our proposed approach for efficient, resource-aware closed-loop learning by exploiting time-series information.
First, we introduce our method for integrating time-series data into the Bayesian optimization (BO) framework through an additional fidelity dimension in the Gaussian process surrogate model.
Subsequently, we detail the development of probabilistic early stopping criteria, which utilize insights from the multi-fidelity surrogate model to assess the promise of ongoing closed-loop experiments and determine whether they should be terminated early to save experimental time.

\subsection{Time-series Informed Bayesian Optimization for MPC Parameter Learning}
Conducting closed-loop experiments is often costly, e.g., in a real-world setting or for computationally expensive (high-fidelity) simulations.
While BO is already sample efficient, the underlying structure of the specific problem is not exploited since BO is a black box approach, leaving potential for further improvement of the sample efficiency.
In the literature, most BO-based approaches for closed-loop learning compress the trajectories observed in a full closed-loop experiment into a single numerical value, namely the value of the performance measure for this specific closed-loop experiment under a specific parameterization.
Doing so, much of the information contained in the closed-loop trajectories is lost, e.g., information about the time-series structure of the closed-loop trajectories and the specific dynamical behavior.
A standard performance measure from the literature on BO for closed-loop learning of controller parameters is given by
\begin{equation}
\label{eq:closed-loop_measure}
    G(\theta) = - \sum_{k=0}^M l_{\text{cl}}(x_k, u_k).
\end{equation}
Here, $M$ is the length of the closed-loop experiment and $l_{\text{cl}} : \mathbb{R}^{n_{\text{x}}} \times \mathbb{R}^{n_{\text{u}}} \to \mathbb{R}, (x_k, u_k) \mapsto l_{\text{cl}}(x_k, u_k)$ is a closed-loop stage cost depending on the closed-loop states $x_k$ and closed-loop control inputs $u_k$.
Both $x_k$ and $u_k$ result from executing the parametrized controller $\pi(\cdot;\theta)$ on the system and therefore depend on the controller parameters $\theta$.
While the above approach is suitable for optimization in a true black box setting, there is additional knowledge available that can be exploited in the case of closed-loop learning.
This specifically concerns the time-series structure of the underlying problem.
We hypothesize that the sample efficiency is enhanced by incorporation of additional evaluations of the closed-loop measure along the time-series.
Specifically, we evaluate the closed-loop measure already at early stages of the closed-loop experiment, i.e., we break up the sum in \eqref{eq:closed-loop_measure} into $L-1$, $L \in \mathbb{N}$ parts\footnote{For simplicity, we choose $L$ discrete fidelity levels uniformly along the closed-loop experiment, i.e., $s = l/L$ with $l \in \{1,\dots,L\}$. This assumption is not restrictive — more general choices of $\mathcal{S}$ (non-uniform grids or even continuous fidelities in $s$) are fully compatible with the proposed approach. We additionally assume $M/L \in \mathbb{N}$, which is also not restrictive.} and include these low-fidelity estimates into the GP surrogate model according to
\begin{equation}
\label{eq:mf_closed-loop_measure}
    \bar G(\theta, s=l/L) = - \sum_{k=0}^{l(M/L)} l_{\text{cl}}(x_k, u_k)
\end{equation}
where $l \in [ 1, \dots, L ]$ is an index for the low-fidelity estimates and, as introduced in Section \ref{sec:trace_aware_bo}, $\bar G(\theta, s=1) = G(\theta)$ holds. 
This approach yields $L-1$ data points from a single closed-loop experiment instead of $1$ when compared to the standard black box BO approach.
Thus, in the specific case of closed-loop learning, we define the target fidelity $s=1$ as consideration of the full length time series.
While the early stages represent only incomplete information, they already provide insights into the quality of the specific closed-loop experiment.
Essentially, we align the fidelity dimension of the GP surrogate with the time axis of the closed-loop experiment and call our approach \textit{time-series-informed Bayesian optimization} (TSI-BO).
Note that in theory, all time-discrete states can be included by choosing $L=M$. However, in practice $L$ can be flexibly chosen to balance information content and computational cost of GP inference, especially when $M$ is large. Consequently, we typically select $L < M$ to reduce the number of fidelity evaluations and thus limit computational effort, while still retaining sufficient temporal resolution of the closed-loop behavior.
Finally, we note that this construction assumes that the closed-loop cost is additive along time, which is satisfied by many common closed-loop performance measures but may require adaptation in settings with non-additive objectives.
The time-series-informed Bayesian optimization approach is illustrated in Figure \ref{fig:approach}, showcasing the alignment of the fidelity dimension $s$ and the time axis of the trajectories observed during a closed-loop experiment, as well as the intermediate performance measure evaluations.

Based on the performance observed early on in a closed-loop experiment, we exploit the GP surrogate model $\hat{\bar{G}}$ to determine whether a closed-loop experiment will have a promising outcome at target fidelity.
To this end, we introduce decision criteria in the next section.

\subsection{Early-Stopping Criteria}
We additionally exploit the surrogate Gaussian process model $\hat{\bar{G}}(\theta, s)$, see Section \ref{sec:trace_aware_bo}, to stop a closed-loop experiment as soon as it seems unpromising.
Specifically, we employ the introduced low-fidelity performance measure evaluations along the time-axis of the closed-loop experiment.
At each evaluation point, we predict the posterior mean $m^+(\theta_n, s=1)$ and posterior standard deviation $\sigma^+(\theta_n, s=1) = \sqrt{k^+(\theta_n, s=1)}$ for the current parameterization $\theta_n$ and at target fidelity $s=1$ using the multi-fidelity GP surrogate.
This way, we exploit information from the data observed so far in the current closed-loop experiment as well as the previous closed-loop experiments -- and the correlations between them as captured in the GP surrogate -- to decide if the current experiment is promising, i.e., if the closed-loop measure at target fidelity $s=1$ is expected to be good enough.

We propose two distinct probabilistic decision criteria, to determine whether a closed-loop experiment is promising or if it should be aborted to save experimental resources.
First, we propose a decision criterion based on the upper confidence bound (UCB) of the current parameterization at target fidelity.
It is defined as
\begin{equation}
\mathcal{E}_{\text{UCB}} : m^+(\theta_n, s=1) + \beta \ \sigma^+(\theta_n, s=1) < G_n^*,
\end{equation}
where $\beta \in \mathbb{R}_{>0}$ is a confidence hyperparameter and $G_n^*$ is the best closed-loop measure value seen so far at target fidelity, i.e.,
\begin{equation}
    G_n^* = \max_{i \in \{1,\dots,n\}} \; \bar{G}(\theta_i, s=1).
\end{equation}
Essentially, the UCB criterion classifies a closed-loop experiment as unpromising as soon as the predicted closed-loop measure for the current experiment is lower than the currently best seen closed-loop performance measure with some confidence encoded by $\beta$.

We propose a second decision criterion based on the expected improvement (EI) of the current parameterization at target fidelity over the best parameterization seen so far.
The criterion is given by
\begin{align*}
\mathcal{E}_{\mathrm{EI}}:\; &(m^+(\theta_n,1) - G_n^*)\,\Phi(Z) + \sigma^+(\theta_n,1)\,\phi(Z) < \tau_{\mathrm{EI}}, \\
Z &= \frac{m^+(\theta_n,1) - G_n^*}{\sigma^+(\theta_n,1)}.
\end{align*}
Here, $\Phi(Z)$ and $\phi(Z)$ are the cumulative distribution function and the probability density function of the standard normal distribution, respectively, the hyperparameter $\tau_{\text{EI}} \in \mathbb{R}_{>0}$ is a threshold value, and $Z$ is the normalized improvement term.

In addition to the criteria based on the GP surrogate, we propose a straightforward convergence criterion based on the closed-loop states, which indicates if a closed-loop run converged to the desired state.
The closed-loop state convergence criterion is given by
\begin{equation}
\label{eqn:state_convergence_criterion}
\mathcal{E}_{\text{C}} : \|x_k - x_\text{d}\| < \epsilon,
\end{equation}
where $\| \cdot \|$ is the two-norm and $\epsilon \in \mathbb{R}_{>0}$ is a convergence threshold.

Combining the concepts of time-series-informed Bayesian optimization and the early stopping criteria, which are tightly linked through the GP surrogate model, the resulting closed-loop tuning procedure is summarized in Algorithm~\ref{alg:tsibo}. We next evaluate the performance of the proposed approach in simulation.

\section{Simulation Studies}
\label{sec:simulation}
We illustrate the effectiveness of the proposed approach in simulation.
After introducing our set-up, we show simulation results for the proposed time-series-informed closed-loop learning approach.
These results are compared against a black box BO baseline for closed-loop learning, focusing on performance metrics and the experimental resources required to achieve convergence.

\begin{algorithm}[b]
\vspace{5mm}
\caption{TSI-BO with Early Stopping}
\label{alg:tsibo}
\begin{algorithmic}[1]
\State Initialize dataset $\mathcal{D} \leftarrow \emptyset$, select initial parameters $\theta_0$
\State Define $\mathcal{K}_{\mathrm{eval}} \coloneqq \{\, k_l = l(M/L)\;|\; l=1,\dots,L \,\}$
\For{$n = 0,1,\dots$}
    \State Execute closed-loop experiment with parameters $\theta_n$
    \For{$k = 0,1,\dots,M$} \Comment{Closed-loop experiment}
        \State Apply $u_k=\pi(x_k;\theta_n)$, observe $(x_{k+1},u_k)$
        \If{$k \in \mathcal{K}_{\mathrm{eval}}$} \Comment{Discrete fidelity level}
            \State $s \gets k/M$; \quad compute $\bar G(\theta_n, s)$
            \State Add $(\theta_n, s, \bar G(\theta_n,s))$ to $\mathcal{D}$
            \State Update multi-fidelity GP surrogate
            \If{$\mathcal{E}_{\mathrm{UCB}}$/$\mathcal{E}_{\mathrm{EI}}$ \textbf{or} $\mathcal{E}_{\mathrm{C}}$ is triggered}
                \State \textbf{break} \Comment{terminate episode early}
            \EndIf
        \EndIf
    \EndFor
    \State Select $\theta_{n+1}$ according to \eqref{eqn:takg_optimization} at $s{=}1$
\EndFor
\end{algorithmic}
\end{algorithm}

\subsection{Simulation Setup}
\begin{figure*}[t]
    \vspace{1mm}
    \centering
    \includegraphics[width=0.99\textwidth]{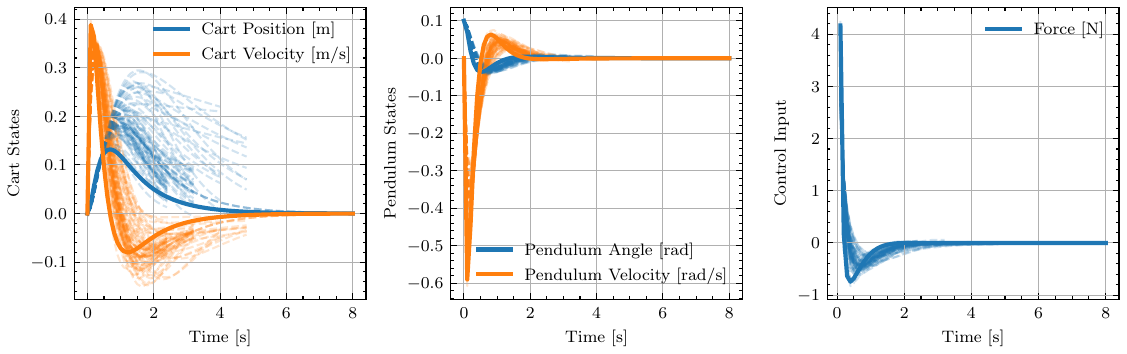}
    \vspace{-4mm}
    \caption{All sampled closed-loop trajectories for a single run of the proposed TSI-BO procedure using EI-based and convergence-based early stopping. Early termination of unpromising or already-converged episodes illustrates the resource savings achieved by the proposed approach.}
    \label{fig:cartpole_trajectories}
    \vspace{-7mm}
\end{figure*}
In our simulation study, we consider the nonlinear cart pole system, a benchmark problem in control theory.
The system consists of a cart moving along a track while balancing an inverted pendulum.
The control input $u$ is the horizontal force applied to the cart.
The control task is to perform a set-point change starting from an initial state $x_0$ and controlling the pendulum to the upright position while placing the cart at the origin.
This corresponds to a desired state $x_\text{d} = \begin{bmatrix} 0, 0, 0, 0 \end{bmatrix}^\intercal$ and a control input $u_\text{d} = 0$.
We choose the closed-loop stage cost as $l_{\text{cl}}(x_k, u_k) = x_k^\intercal Q_{\text{BO}} x_k$, where $Q_{\text{BO}} \in \mathbb{R}^{4 \times 4}$ is a constant weight matrix.
While the chosen closed-loop stage cost $l_{\text{cl}}$ looks structurally similar to the standard quadratic MPC stage cost $l_\theta$, it is not parameterized and considered over the full length $M$ of a closed-loop experiment, rather than the short MPC prediction horizon $N$, i.e., $M \gg N$.
We choose $M = 80$ and $N=20$.
Specifically, the MPC stage cost used inside the controller is parameterized as
\begin{equation}
    l_\theta(x_k,u_k) = x_k^\top Q(\theta) x_k + u_k^\top R(\theta) u_k,
\end{equation}
where we parameterize only the main diagonal of the state weight matrix $Q(\theta) = \mathrm{diag}(\theta_1,\dots,\theta_4)$ and the scalar input weight $R(\theta)=\theta_5$. No other part of the MPC formulation is parameterized. Thus, the BO is performed over $n_\mathrm{p} = 5$ controller parameters in total.

We introduce $L = 10$ evenly distributed performance measure evaluation points in the closed-loop experiment.
At each evaluation point, we incorporate the so-far observed closed-loop performance into the multi-fidelity BO procedure.
The cost defined in the trace-aware knowledge gradient acquisition function is not used in our work.
Instead, we employ early stopping criteria to save experimental resources by terminating unpromising experiments based on the predicted high-fidelity outcomes.
For the Gaussian process surrogate model, we use an exponential decay kernel for the fidelity dimension \cite{wu2020practical}, allowing us to effectively capture correlations across the fidelity levels. This choice is well suited for step-like episodes with monotone decay in cost. However, alternative kernels can be used for non-monotone tasks. Additionally, we employ a Matérn $5/2$ kernel for the parameter dimensions.

\subsection{Simulation Results}
The closed-loop trajectories for one exemplary run are shown in Figure \ref{fig:cartpole_trajectories}.
It is apparent that many closed-loop experiments are aborted, thus, saving valuable experimental resources.
This is particularly visible for trajectories that show a high overshoot in the cart position.
Since the overshoot leads to a bad closed-loop performance, the experiments are aborted early on, according to the proposed early stopping criteria $\mathcal{E}_{\text{UCB}}$ and $\mathcal{E}_{\text{EI}}$.

\begin{figure}
    \vspace{1mm}
    \begin{minipage}[b!]{0.47\textwidth}
        \includegraphics[width=\textwidth]{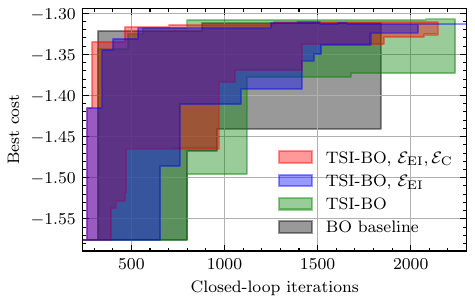}
    \end{minipage}
    \begin{minipage}[b!]{0.47\textwidth}
        \vspace{-9mm}
        \includegraphics[width=\textwidth]{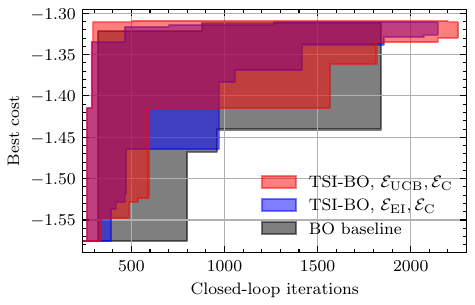}
    \end{minipage}
    \vspace{-2mm}
    \caption{Comparison of best-so-far cost across 10 independent runs. The top panel shows ablations of the proposed TSI-BO method, including the BO baseline, TSI-BO with $\mathcal{E}_\mathrm{EI}$ and $\mathcal{E}_\mathrm{C}$, TSI-BO with $\mathcal{E}_\mathrm{EI}$ only, and TSI-BO without early stopping. The bottom panel compares the EI- and UCB-based early stopping criteria against the same BO baseline. Shaded areas indicate the minimum--maximum range across runs, and the horizontal axis denotes the number of closed-loop iterations (i.e., experimental resources).}
    \label{fig:cartpole_regret}
    \vspace{-8mm}
\end{figure}

In Figure~\ref{fig:cartpole_regret}, we report the best-so-far cost versus closed-loop iterations, where the latter represent the required experimental resources. 
The shaded envelopes span the minimum--maximum range across $10$ independent runs.
For the TSI-BO variants with early stopping, some runs terminate earlier than others once their experimental budget is exhausted. 
As a consequence, the maximum of the envelope may decrease at certain points, since no further best-cost updates occur in those terminated runs.

In Figure~\ref{fig:cartpole_regret} (top), we show ablations of the proposed TSI-BO method, where all TSI-BO variants outperform the BO baseline. 
Using TSI-BO without early stopping already accelerates best cost convergence, indicating that aligning the fidelity dimension with closed-loop time is beneficial on its own. 
Adding the EI-based early stopping criterion further speeds up convergence, while combining EI with the convergence rule~$\mathcal{E}_\mathrm{C}$ yields the strongest resource savings by truncating unpromising or already-converged episodes early, thus improving learning efficiency without requiring additional experiments.
Nevertheless, the simple convergence criterion~$\mathcal{E}_\mathrm{C}$ alone does not substantially improve convergence speed beyond the EI-based stopping rule, indicating that most of the gains arise from the probabilistic early-stopping mechanism rather than from state-based convergence detection.

In Figure~\ref{fig:cartpole_regret} (bottom), we compare the EI- and UCB-based early stopping criteria, both combined with~$\mathcal{E}_\mathrm{C}$, against the same BO baseline. 
Both approaches dominate the baseline across nearly the entire experimental budget and reduce regret much faster, 
with only small differences between EI and UCB remaining within the min--max envelopes. 
This demonstrates that both criteria achieve similar convergence behavior and learning efficiency in practice.

Beyond faster convergence, the TSI-BO variants also achieve a better final cost compared to the black-box BO baseline. 
Across all ablations, the envelopes of the TSI-BO methods remain above those of the baseline throughout the optimization process. 
Among the variants, the combination of the EI-based early stopping rule with the convergence criterion~$\mathcal{E}_\mathrm{C}$ yields the most favorable final cost, closely followed by TSI-BO with only~$\mathcal{E}_\mathrm{EI}$. 
The version without early stopping still outperforms the baseline, confirming that exploiting the time-series fidelity alone already enhances optimization efficiency. 
In comparison, the UCB-based stopping rule achieves a comparable final performance, with small differences observable across runs, which might be attributed to its different truncation behavior.

Since the plots show the full range (minimum--maximum) across runs rather than mean and variance, the narrower envelopes of the TSI-BO curves indicate lower variability and more consistent performance across independent runs. 
In contrast, the baseline exhibits both lower and more fluctuating best-cost values, reflecting its greater sensitivity to random initialization and limited data efficiency.

Finally, the smoother appearance of the baseline curve arises from its lower sampling density: only one high-fidelity cost is obtained per full episode. 
In contrast, time-series-informed Bayesian optimization continuously updates the surrogate with partial-fidelity data during each experiment, resulting in a denser regret trace. 
This difference further underlines the resource inefficiency of the baseline, which discards valuable intermediate trajectory information that time-series-informed Bayesian optimization effectively exploits to accelerate learning.

\section{Conclusion}
\label{sec:conclusion}
This work presented a time-series-informed Bayesian optimization framework for resource-efficient closed-loop controller parameter tuning. By aligning the multi-fidelity surrogate with the closed-loop time axis and incorporating intermediate partial-episode evaluations, the method exploits temporal structure already available during an episode. Probabilistic early stopping rules further reduce unnecessary full experiment execution by terminating unpromising parameterizations early based on surrogate belief.
Simulation results show that the proposed approach achieves comparable closed-loop performance using roughly half the experimental resources required by standard black-box BO, and yields better final performance under equal resource budgets.
This resource efficiency underscores the potential of exploiting inherent structural information in closed-loop learning tasks to enhance both performance and resource utilization.
Future work will explore extensions of time-series-informed Bayesian optimization towards look-ahead Bayesian optimization methods, e.g., \cite{paulson2022efficient}, to enable sampling of multiple parameterizations within a single closed-loop experiment, further improving resource efficiency and convergence speed.

\vspace{-2mm}
\bibliographystyle{ieeetr}
\bibliography{bibliography}

\end{document}